\documentstyle [12pt]{article}
\begin{document}
\newcommand{\sptwo}{1.8}
\newcommand{\spone}{0.9}
\newcommand{\doublespace}{\edef\baselinestretch{\sptwo}\Large
\normalsize}
\newcommand{\singlespace}{\edef\baselinestretch{\spone}\Large
\normalsize}
\parindent=2.5em
\thispagestyle{empty}
\baselineskip=22pt

\hspace*{\fill} {PURD-TH-92-11} \\
\hspace*{\fill} {October 1992}

\begin{center}

{}~\\
\LARGE{Mass of the Lightest Higgs Boson in
the Minimal Supersymmetric Standard Model
with an Additional Singlet}
{}~\\
{}~\\
\large{~\\
W. T. A. ter Veldhuis\\
{}~\\
Department of Physics\\
Purdue University\\
West Lafayette, IN 47907, U.S.A.\\
{}~\\
{}~\\
{}~\\}

\end{center}

\doublespace

\begin{abstract}

An upperbound on the mass of the
lightest neutral scalar Higgs boson is calculated in an extended
version of the minimal supersymmetric standard model that contains
an additional Higgs singlet. We integrate the
renormalization group
equations of the model, and impose low energy boundary conditions
consistent with present experimental results, and ultra-violet conditions
following from triviality. Radiative corrections induced by a
large top quark Yukawa coupling are included in our analysis,
and we find the allowed values for the mass of
the Higgs boson as a function of the mass of the top
quark. Typically, for a top quark mass $m_t=150\ GeV$,
the upper bound on the Higgs boson mass is about $25 \ GeV$ higher than
in the minimal model.

\end{abstract}

\newpage

\begin{center}
\section*{}
\end{center}

The predictions of the standard model are in excellent agreement
with current experimental results. However, many questions
remain unanswered, and in particular the
situation in the Higgs sector is theoretically unsatisfactory.
The standard model is not natural, in the sense that it does not
explain why the electroweak scale is so tiny compared to the Planck
scale.
Moreover, the presence
of an elementary scalar leads to quadratic divergencies, and as a
consequence the scalar mass parameter has to be
fine tuned in all orders of
perturbation theory.
Supersymmetry
provides an attractive solution to this technical aspect of
the naturalness problem.
It is well known that the minimal
supersymmetric standard model
predicts a neutral Higgs boson with a mass below $M_Z$ at tree
level. Recently, it was recognized that radiative corrections due to
a large top Yukawa coupling can give a significant contribution to
the mass, raising its limit, typically by $25 \ GeV$ \cite{Ell,Yan2,Dia}.
This upper bound can be regarded as
a very attractive feature, making the model subject
to experimental verification. However as the experimental lower bound
on the mass of a neutral Higgs boson increases,
the question whether it is possible to extend the minimal
model in a simple way
to raise the upper bound on the Higgs boson mass becomes more
relevant.
In the present letter we therefore calculate
an upper bound on the mass of the lightest
neutral Higgs boson in the minimal supersymmetric standard model with
an additional Higgs singlet. This extended model has been
studied in some detail \cite{Ell2,Dre,Gun}, however, as
the top quark was thought to be much lighter at the time, radiative
corrections due to a large top Yukawa coupling were not included.
The implications of unitarity constraints at the unification scale on the mass
of the lightest neutral Higgs boson were studied using renormalization
group techniques
by Durand and Lopez \cite{Dur}.
We follow a similar tack, but the bound we set on the mass of the
lightest neutral scalar Higgs
boson follows from the requirement that perturbation theory
remains valid up to a cut-off scale $\Lambda$. Moreover, we include
the above mentioned radiative corrections in our analysis.
Typically, we take take a supersymmetry breaking scale $M_{SUSY}=10^3\ GeV$
and the cut-off scale $\Lambda=10^{16}\ GeV$, although we study the
mass of the Higgs boson as a function of both $M_{SUSY}$ and $\Lambda$.
Espinosa and Quiros \cite{Esp}, in contrast,
calculated bounds on the Higgs boson
mass by requiring perturbation theory to be valid below $M_{SUSY}$.
Our approach is  analogous to the calculation of
the familiar triviality bound in the regular standard model \cite{Hil,Hil2}.
An important role is played by a so called infrared quasi-fixed point,
which gives rise to strong limits on the parameter
space at low energy, insensitive to the precise value
of the cut-off $\Lambda$.
Recently, Ellwanger et al. \cite{Ulr} used a crude method to
obtain an estimate for the range of all Higgs masses in this
extended model,
taking into account the heavy top quark radiative corrections.

To be concrete, let us
consider the Higgs sector of the softly broken supersymmetric
standard model with an additional singlet field \cite{Ell2,Dre}:
\begin{eqnarray}
\Gamma & = & \int dV \left[
(1 + \hat{m}_T^2\theta^2\overline{\theta}^2) H_T \overline{H}_T +
(1 + \hat{m}_B^2\theta^2\overline{\theta}^2) \overline{H}_B H_B +
(1 + \hat{\mu}^2\theta^2\overline{\theta}^2) \overline{S} S
\right] +  \nonumber   \\
 & & \int dS  \left[
\xi S +
\frac{\mu}{2}(1 + b_s \theta^2) S^2  +
\frac{\lambda_s}{3}(1 + a_s \theta^2) S^3 \right] + h.c. + \nonumber \\
 & & \int dS  \left[ m (1 + b \theta^2) H_T H_B
+ g (1 + a \theta^2)S H_T H_B \right] + h.c..
\end{eqnarray}
Here $H_T$ and $H_B$ are $SU(2)$ doublets, $S$ is the additional
singlet, and we have included general soft breaking terms \cite{Gir}.
It is interesting to
note that, in contrast with the minimal supersymmetric model, it is
possible to induce the correct electroweak symmetry breaking with only
dimensionless supersymmetric coupling constants (i.e.
$\xi=0$ and  $\mu=m=0$ ) as discussed by Nilles \cite{Nil}.
However, here we will
analyze the more general case.

The $SU(2) \otimes U(1)_Y$ symmetry is spontaneously broken into $U(1)_{EM}$,
because the scalar components of the Higgs superfields develop
vacuum expectation values:
\begin{equation}
<A_B>= \frac{1}{\sqrt{2}}
\left(
\begin{array}{c}
v_B \\
0
\end{array}
\right)
,\ \
<A_T>= \frac{1}{\sqrt{2}} \left( v_T,0 \right)
,\ \
<A_S>= \frac{1}{\sqrt{2}} v.
\end{equation}
We will always assume that the coupling
constants are
real. As was shown by Romao \cite{rom},
it then necessarily follows that the vacuum expectation values of
all the Higgs fields are real and the model does not exhibit explicit
CP violation. In fact, the weakness of the observed CP violation gives
a physical motivation for this assumption.

The particle spectrum of this model contains, apart from the three
Goldstone bosons which give rise to
the weak gauge boson masses,
three neutral scalar, two pseudoscalar, and two
charged Higgs bosons. Their mass structure has been studied
by Drees \cite{Dre} and
the symmetric mass matrix for the neutral scalars in the basis
\mbox{
$\frac{1}{\sqrt{2}}Re (A_{T1})$,
$\frac{1}{\sqrt{2}}Re (A_{B1})$,
$\frac{1}{\sqrt{2}}Re (A_S)$ }
takes the form:
\newcommand{\Cone}{\mbox{ $\left( \lambda_s g \frac{v^2}{2} +
ga \frac{v}{\sqrt{2}} +\xi g +\mu \frac{v}{\sqrt{2}} +mb \right)$ } }
\newcommand{\Cfour}{\mbox{ $ \left( 2 \lambda_s g v + g a \sqrt{2}
+\mu g \sqrt{2} \right) $
} }
\newcommand{\Cfive}{\mbox{ $ \left( 2 g^2 v +2 m g \sqrt{2} \right) $ } }
\newcommand{\Csix}{\mbox{ $
- g \frac{M_W^2}{g_2^2} \left( \sin(2\beta)
(a + \mu)
+ 2m \right) \frac{\sqrt{2}}{v} +
2 \lambda_s^2 v^2 +
\lambda_s \frac{v}{\sqrt{2}} (a_s + 3 \mu)-
\frac{\sqrt{2}}{v}\mu \xi
$ } }
\begin{equation}
\left( M_S^2 \right) =
\left(
\begin{array}{ccc}
-C_1 \tan{\beta} + M_Z^2 \cos^2{\beta}&
C_1 + 2 \left( g^2 - \frac{1}{4} g_1^2 - \frac{1}{4} g_2^2
\right) \frac{v_B v_T}{2} &
M_W \left( C_5 \cos{\beta} + C_4 \sin{\beta} \right) \\
. & -C_1 \cot{\beta} + M_Z^2 \sin^2{\beta}&
M_W \left( C_5 \sin{\beta} + C_4 \cos{\beta} \right) \\
. & . & C_6
\end{array}
\right). \label{eq:mat}
\end{equation}
Here
\begin{eqnarray}
C_1 & = & \Cone,  \nonumber \\
C_4 & = & \Cfour, \nonumber \\
C_5 & = & \Cfive, \nonumber \\
C_6 & = & \Csix.  \nonumber
\end{eqnarray}

We will consider the case in which there is only one light Higgs boson.
This situation is realized if $\left| C_1 \right| >>M_Z^2$, and the
lightest scalar mass eigenstate is approximately,
to order $(\frac{M_Z^2}{C_1})$, \mbox{
$\frac{1}{\sqrt{2}} Re \left( \cos \beta A_T + \sin \beta \overline{A}_B
\right)$ }
with corresponding mass:
\begin{equation}
M_H^2 = M_Z^2
\left( {\cos}^2 (2 \beta) + 2 \frac{g^2}{g_1^2 + g_2^2} {\sin}^2 (2 \beta)
\right). \label{eq:hm}
\end{equation}
Even in the case that the condition
$\left| C_1 \right| >>M_Z^2$ is not fulfilled,
equation(\ref{eq:hm}) still provides a useful upper bound on the
mass of the lightest neutral Higgs boson. This can easily be seen
by diagonalizing the two by two submatrix in the left hand top
corner of eq.(\ref{eq:mat}), and calculating the minimum of the lowest
eigenvalue
as a function of $C_1$.

Let us analyze expression (\ref{eq:hm}) as a function of $\beta$.
One has to distinguish two cases.
\newcommand{\x}{\mbox{$2 \frac{g}{g_1^2+g_2^2}$}}
If $\x <1$, then $M_H$ has a maximum for $\beta =0$. This maximum value
is independent of $g$ and  equal to the maximum value
in the minimal supersymmetric standard model. On the other hand,
if $\x >1$, then $M_H$ reaches its maximum value for $\beta =\pi/4$.
In the latter case $g$ determines
the maximum value of $M_H$, and the situation
is considerably different from the minimal supersymmetric standard model.

Now that the spectrum of the Higgs sector at tree level has been
discussed, we will proceed by including one loop quantum corrections in
a renormalization group analysis of the mass of the lightest Higgs
boson. We assume there is no new physics between
the cut-off scale $\Lambda$ and the soft supersymmetry breaking
scale $M_{SUSY}$. The existence of this so called desert enables us to
use the renormalization group equations of the supersymmetric
model to relate the couplings at the cut-off scale $\Lambda$ to their
values at the supersymmetry  breaking scale $M_{SUSY}$.
The requirement of consistency of perturbation theory puts a constraint
on the value of $g(M_{SUSY})$, because the structure of
the renormalization group equations causes the Higgs self couplings
to become singular at an energy scale $\mu<\Lambda$ if their low
energy values are too big.
We therefore  introduce a new variable $\tan(\alpha)$, and
impose high energy boundary conditions on $\lambda_s$ and $g$
by requiring that at least one of them becomes non-perturbative at the
cut-off scale $\mu=\Lambda$:
\begin{equation}
\tan \alpha = \frac{\lambda_s(\Lambda)}{g(\Lambda)}, \ \ \ with \ \ \
g^2(\Lambda)+\lambda_s^2(\Lambda)=100. \label{eq:big}
\end{equation}
The choice of the right hand side of equation(\ref{eq:big}) is quite
arbitrary, as long as the number is large, since the nature
of the renormalization group equation is such that solutions
which are rather far apart at high energy approach each other
at low energy. Defining the scaling variable $t=\ln (\frac{\mu}{M_Z})$,
the evolution of the coupling constants between
$\Lambda$ and $M_{SUSY}$ is
given by their one loop renormalization group equations \cite{Der}:
\begin{equation}
(4 \pi)^2 \frac{d g^2}{dt} = \left( 4 \lambda_s^2 + 8 g^2
+6 h_t^2 -2 g_1^2 -6 g_2^2 \right) g^2 , \label{rg:rgg}
\end{equation}
\begin{equation}
(4 \pi)^2 \frac{d \lambda_s^2}{dt} = \left( 12 \lambda_s^2 + 12 g^2 \right)
\lambda_s^2,
\end{equation}
\begin{equation}
(4 \pi)^2 \frac{d h_t^2}{dt} = \left( 2 g^2 +12 h_t^2
-\frac{26}{9} g_1^2 -6 g_2^2 - \frac{32}{3}g_3^2 \right) h_t^2,
\end{equation}
\begin{equation}
(4 \pi)^2 \frac{d g_i^2}{dt} = {\beta}_i g_i^4,
\end{equation}
with $\beta_1=22$, $\beta_2=2$ and
$\beta_3=-6$.
In the case that only one Higgs boson has mass below $M_{SUSY}$,
the lightest neutral scalar state and
the Goldstone bosons will form
the doublet that is almost a mass eigenstate in the following
way:
\begin{eqnarray}
\overline{H}_H&=&-\sin (\beta) A_T +\cos (\beta) \overline{A}_B, \nonumber \\
H_L&=&\cos (\beta) A_T +\sin (\beta) \overline{A}_B.
\end{eqnarray}
We assume that the heavy Higgs doublet $H_H$ decouples at $M_{SUSY}$,
and the theory below $M_{SUSY}$ is equivalent to the regular standard model,
with Higgs potential:
\begin{equation}
V = m_0^2 \overline{H}_L H_L + \lambda (\overline{H}_L H_L)^2.
\end{equation}
This approximation makes it possible
to relate the coupling constants of the standard model to those
of the supersymmetric model at $M_{SUSY}$. To be concrete, we
find the following boundary conditions at $M_{SUSY}$:
\begin{equation}
\lambda = \frac{1}{8} \left( g_1^2 + g_2^2 \right)
\left( {\cos}^2 (2 \beta) + 2 \frac{g^2}
{g_1^2 + g_2^2} {\sin}^2 (2 \beta)
\right), \label{bc:lambda} \nonumber
\end{equation}
\begin{equation}
h_t' = h_t \cos (\beta). \label{bc:ht}
\end{equation}
Here $h_t'$ is the top Yukawa coupling in the standard model.
In addition, below the supersymmetry
breaking scale $M_{SUSY}$, the superpartners
decouple and
do not contribute to the renormalization group equations any more.
In order to take this into account, we utilize the
renormalization group equations of the standard model to run
the coupling constants down from  $M_{SUSY}$ to the electroweak scale.
To one loop order these renormalization group equations are \cite{Bar}:
\begin{equation}
(4 \pi)^2 \frac{d \lambda}{dt} =
24 {\lambda}^2 + \left( 12 h_t'^2 -3 g_1^2 -9 g_2^2 \right) \lambda
+ \frac{3}{8} g_1^4 + \frac{3}{4} g_1^2 g_2^2 + \frac{9}{8} g_2^4
-6 h_t'^4, \label{eq:rglambda}
\end{equation}
\begin{equation}
(4 \pi)^2 \frac{d h_t'^2}{dt} =
9 h_t'^4 - \left( \frac{17}{6} g_1^2 + \frac{9}{2} g_2^2  + 16 g_3^2 \right)
h_t'^2,
\end{equation}
\begin{equation}
(4 \pi)^2 \frac{d g_i^2}{dt} = {\beta}_i g_i^4,
\end{equation}
with $\beta_1=\frac{41}{3}$, $\beta_2=-\frac{19}{3}$ and
$\beta_3=-14$.
We thus include leading log radiative corrections
proportional to $\ln(\frac{M_{SUSY}}{M_Z})$,
in particular those caused by
a large top Yukawa coupling $h_t$,
but neglect finite corrections.
Current experimental results lead us to impose the following
low energy boundary conditions on the gauge coupling constants:
\mbox{$g_1^2(M_Z) = 0.1282$},
\mbox{$g_2^2(M_Z) = 0.4222$}
and \mbox{$g_3^2(M_Z) = 1.445$}.
To complete the set of input parameters, we take $M_Z=91.0\ GeV$.
The vacuum expectation value $v_0$ of the light Higgs boson is then fixed
to be:
\begin{equation}
v_0 =  2 \frac{M_Z}{\sqrt{g_1^2 + g_2^2}},
\end{equation}
and $m_t$ and $m_H$ are defined as follows:
\begin{equation}
m_H^2 = 2 \lambda(m_H) v_0^2,\label{eq:defmh}
\end{equation}
\begin{equation}
m_t^2 = \frac{v_0^2}{2} h_t'^2(m_t).
\end{equation}

The mass of the lightest neutral Higgs boson is enhanced by radiative
corrections as well as by the second term in eq.(\ref{eq:hm}).
An estimate of the magnitude of the radiative corrections
can be obtained in the following way \cite{Yan}. If the top
quark is heavy, the $h_t'^4$ term in eq.(\ref{eq:rglambda}) will
dominate. Keeping only this term and assuming $h_t'$ to be
constant, the renormalization group equation can be solved, and
one obtains:

\begin{equation}
(4 \pi)^2 \lambda(t)= (4 \pi)^2 \lambda(t_{SUSY}) + 6 h_t'^4 (t_{SUSY}-t).
\end{equation}
This leads to a positive contribution to $m_H^2$:

\begin{equation}
\delta m_H^2 \approx \frac{48}{\left(4 \pi \right)^2} \frac{m_t^4}{v_0^2}
\ln{\left(\frac{M_{SUSY}}{m_H}\right)}.
\end{equation}
Note however, the situation here differs from the
minimal supersymmetric standard model, because $h_t$ affects the
running of $g$ above $M_{SUSY}$. It can be easily seen from
equation(\ref{bc:ht}) that $h_t$ becomes larger if $\beta$ increases
for a fixed value of the top quark mass $m_t$. Moreover, according
to equation(\ref{rg:rgg}), in that case
the value of $g(M_{SUSY})$  decreases
for a given value of $g(\Lambda)$. However, it follows from
boundary condition(\ref{bc:lambda}) that if $\x>1$, $\lambda(M_{SUSY})$
is maximal for a given value of $g(M_{SUSY})$ for $\beta=\frac{\pi}{4}$.
We therefore expect that the mass of the lightest neutral scalar
Higgs boson reaches
its maximum for a value of $\beta$ somewhere in between
$0$ and $\frac{\pi}{4}$ for sufficiently small $\tan(\alpha)$.

Before introducing our results, we now outline our computational procedure.
We choose values for the free  parameters $\tan \alpha$,
$\tan \beta$, $\Lambda$ and $M_{SUSY}$, and we pick a value
for the top quark mass $m_t$ in the currently expected range.
Subsequently, we
numerically integrate the system of differential equations and
find a solution that satisfies all boundary conditions. The mass
of the lightest neutral Higgs boson is then extracted
using equation(\ref{eq:defmh}). Since we neglect the
bottom quark Yukawa coupling, our results are valid if $h_t>>h_b$,
which means $\tan(\beta)>>\frac{m_b}{m_t}\approx0.03$.

In figure 1 we plot the mass of the lightest neutral Higgs boson as a function
of $\tan \beta$. Motivated by grand unification, we choose
a cut-off scale $\Lambda= 10^{16} \ GeV$ and $M_{SUSY}= 10^3 \ GeV$.
For small values of $\tan (\beta)$, the first term in
equation(\ref{bc:lambda}) dominates. This term is independent of
$\lambda_s$ and $g$, and so for all values of $\tan(\alpha)$ we find
that the mass of the lightest neutral Higgs boson is equal to
the mass in the minimal supersymmetric standard model
enhanced by the heavy top quark radiative corrections. However,
if we follow the  Higgs mass towards larger values of $\tan(\beta)$,
we observe that for small values of $\tan(\alpha)$ the mass reaches a maximum
of $130 \ GeV$ for $\tan(\beta)=0.6$. On the other hand, for large values
of $\tan(\alpha)$, the Higgs mass decreases monotonously as a function
of $\tan(\beta)$. In fact, in the limit $\tan(\alpha) \rightarrow \infty$ the
singlet decouples, and we obtain the minimal supersymmetric standard model.

A cutoff scale of $10^{16} \ GeV$ is attractive because
the flow of the renormalization group
in combination with the low energy boundary conditions forces the trajectories
of the three gauge couplings to intersect at one point.
However, we do not limit ourselves to this case,
and in figure 2
we plot  the triviality bound on the
mass of the lightest neutral scalar Higgs boson as a function of the
cut-off scale $\Lambda$.
Again we chose $m_t = 150 \ GeV$, $M_{SUSY}=1 \ TeV$ and we took $\tan
(\beta)=0.6$
and $\tan (\alpha)=0.1$, since the Higgs boson mass reaches
its maximum for these values. As the cut-off scale $\Lambda$ is lowered,
the upper bound on the lightest neutral Higgs mass increases,
and in the limit
$\Lambda \rightarrow M_{SUSY}$ the upper bound on the lightest Higgs boson
approaches $M_H=800 \ GeV$, the triviality bound in the regular
standard model.

In figure 3 we show a triviality plot of $m_H$ versus $m_t$ for a
cut-off scale $\Lambda = 10^{16} \ GeV$ and a supersymmetry breaking
scale $M_{SUSY}= 1 \ TeV$.
The boundary of the enclosed
area indicates that either $\lambda_s$, $g$ or $h_t$ becomes
non-perturbative for $\mu \leq \Lambda$ for all values of $\tan \beta$
and $\tan \alpha$. In other words, there exist finite values
of the coupling constants $\lambda_s(\Lambda)$, $g(\Lambda)$ and
$h_t(\Lambda)$ that give combinations of the top quark mass $m_t$
and the lightest neutral Higgs boson mass $M_H$ within this region.
The dashed line indicates combinations of $m_t$ and $M_H$ that
can only be reached if $h_t(\Lambda)$ becomes singular.
The dotted lines below the solid line show $M_H$ as a function
of $m_t$, for $\tan(\alpha)=10.0$ and various values of
$\tan(\beta)$. As we have discussed before, in the limit of large
$\tan(\alpha)$ we obtain the minimal supersymmetric standard model.
Hence, points on the dashed line represent the predictions of the
supersymmetric top quark condensate model \cite{Cla,Sas,Car}.
In particular,  the fact that
$h_t(\Lambda) \rightarrow \infty$ corresponds exactly  to
the compositeness condition in
that model. The solid line shows the relation between $M_H$ and
$m_t$ for $\tan(\alpha)=10.0$ and $\tan(\beta)= 0.1$. Consequently,
the dashed line and the solid line encompass the area of allowed $m_t$
and $M_H$ in the minimal supersymmetric standard model. As a matter of
fact, the solid line gives the sum of $M_Z$ and the radiative corrections
to the Higgs mass as a function of $m_t$ in the minimal
supersymmetric model.
For small values of $m_t$, the maximum value of $M_H$ is reached for
$\beta=\frac{\pi}{4}$, but as the top mass increases, $M_H$ is maximal
for lower values of $\beta$. Indeed, for very large values of
$m_t$ the highest value for $M_H$ is obtained for $\beta=0$.
The dotted lines above
the solid line show $M_H$ as a function of
$m_t$ for $\tan(\alpha)=0.1$ and various values of $\tan(\beta)$.
The envelop of these dotted lines gives the
maximum of $M_H$ as a function of $m_t$ in the extended model.

Finally, we show in figure 4 how the mass of the lightest neutral
Higgs boson depends on the SUSY breaking scale $M_{SUSY}$ for
various values of $\tan \beta$ and $\tan \alpha$.
It is clear that the upper bound on the lightest neutral Higgs boson
increases as $m_{SUSY}$ becomes larger, since the radiative corrections
become stronger.

In conclusion, we have studied the mass of the lightest neutral
scalar Higgs boson
in an extended version of the minimal supersymmetric standard model,
with an additional  Higgs singlet.
We have included radiative corrections in the leading log approximation.
Since we have assumed hard decoupling of the super-partners, we have
neglected threshold effects. Moreover, we have assumed that one
Higgs boson has mass below $M_{SUSY}$,
and that the heavy doublet and singlet decouple at $M_{SUSY}$.
However, even if this is not the case realized in nature, our
upper bounds remain valid. At tree level it is obvious that
the lightest neutral scalar mass eigenstate becomes lighter if other
Higgs bosons do not decouple. Quantum effects would manifest themselves
by a different running of the coupling constants below $M_{SUSY}$,
but the running of the top Yukawa coupling, the main source of
radiative corrections, would not be affected significantly. We have
checked this explicitly by running the coupling constants down
with the appropriate renormalization group equations for a
model with two Higgs doublets and one Higgs singlet below $M_{SUSY}$.
The upper bound on the lightest neutral scalar Higgs calculated
in this way is virtually identical to the mass we calculate
in our model with only one light Higgs doublet.
As a consequence the upper bounds we calculate are
independent of the details of the soft supersymmetry breaking parameters.
Furthermore we
have ignored the bottom quark Yukawa coupling. In light of
the large mass difference between the top and bottom quark, this
assumption seems very reasonable. The results of our calculations
show that, in particular for a low top quark mass, the lightest
neutral scalar Higgs boson can be significantly heavier than
in the heavy top enhanced minimal supersymmetric standard model.
To be more specific,
for $m_t=100 \ GeV$, $M_H$ can be as much as $55 \ GeV$ heavier than
in the minimal model, and for $m_t=150 \ GeV$, a value more likely,
the upper bound on $M_H$ is about $25 \ GeV$ higher.

\vspace{.25 in}

\noindent
{\bf Acknowledgement}

\noindent
The author would like to thank Tom E. Clark for useful suggestions and
interesting comments.

\newpage

{\bf \section*{FIGURE CAPTIONS}}

\vspace{0.25 in}

{\bf Fig. 1.} Mass of the lightest neutral Higgs boson $M_H$ as a function of
$\tan(\beta)$, for various values of $\tan(\alpha)$: $\tan(\alpha)=0.1$ (solid
line), $\tan(\alpha)=1.0$ (dotted line), $\tan(\alpha)=3.0$ (dashed line),
$\tan(\alpha)=6.0$ (dot-dashed line) and $\tan(\alpha)=10.0$
(dot-dot-dashed line).

\vspace{0.25 in}

{\bf Fig. 2.} Upper bound on the mass of the lightest neutral Higgs boson
as a function of the cut-off scale $\Lambda$, for a top quark mass
$m_t=150 \ GeV$ and a supersymmetry breaking scale $M_{SUSY}=10^3 \ GeV$.

\vspace{0.25 in}

{\bf Fig. 3.} Triviality diagram, indicating possible values of the
top quark mass $m_t$ and lightest neutral Higgs boson mass $M_H$
consistent with perturbation theory for a cut-off scale
$\Lambda= 10^{16} \ GeV$ and a supersymmetry breaking scale
$M_{SUSY}= 10^3 \ GeV$.

\vspace{0.25 in}

{\bf Fig. 4.} Upper bound on the mass of the lightest neutral Higgs boson $M_H$
as a function of the SUSY breaking scale $M_{SUSY}$, for a top quark mass
$m_t=150 \ GeV$, a cut-off scale $\Lambda=10^{16} \ GeV$ and various
combinations of $\tan(\alpha)$ and $\tan(\beta)$:
$\tan(\alpha)=0.1$ and $\tan(\beta)=0.6$ (solid line),
$\tan(\alpha)=0.1$ and $\tan(\beta)=0.1$ (dashed line),
$\tan(\alpha)=10.0$ and $\tan(\beta)=0.6$ (dotted line),
and finally $\tan(\alpha)=10.0$ and $\tan(\beta)=0.1$ (dot-dashed line).

\end{document}